\documentclass{article}
\usepackage{ismir,amsmath,cite,url}
\usepackage{graphicx}
\usepackage{color}
\usepackage[pagebackref,breaklinks,colorlinks,bookmarks=false]{hyperref}
\usepackage{url}
\usepackage{caption}
\usepackage{subcaption}
\usepackage{xcolor}
\definecolor{pinegreen}{rgb}{0.0, 0.47, 0.44}
\definecolor{oceanblue}{rgb}{0.0, 0.4, 0.7}
\definecolor{cornellred}{rgb}{0.7, 0.11, 0.11}
\definecolor{cadmiumgreen}{rgb}{0.0, 0.42, 0.24}
\definecolor{spirodiscoball}{rgb}{0.06, 0.75, 0.99}
\hypersetup{
  linkcolor = pinegreen,
  citecolor  = oceanblue,
  colorlinks = true,
}

\title{Music2Video: Automatic Generation of Music Video with fusion of audio and text}

\threeauthors
  {Yoonjeon Kim} {KAIST \\ {\tt yoonkim313@kaist.ac.kr}}
  {Joel Jang} {KAIST \\ {\tt wkddydpf@kaist.ac.kr}}
  {Sumin Shin} {KAIST \\ {\tt sym807@kaist.ac.kr}}

\begin{document}

\maketitle
\begin{abstract}
% The abstract should be placed at the top left column and should contain about 150-200 words.
Creation of images using generative adversarial networks has been widely adapted into multi-modal regime with the advent of multi-modal representation models pre-trained on large corpus. Various modalities sharing a common representation space could be utilized to guide the generative models to create images from text or even from audio source. Departing from the previous methods that solely rely on either text or audio, we exploit the expressiveness of both modality. Based on the fusion of text and audio, we create video whose content is consistent with the distinct modalities that are provided. A simple approach to automatically segment the video into variable length intervals and maintain time consistency in generated video is part of our method. Our proposed framework for generating music video shows promising results in application level where users can interactively feed in music source and text source to create artistic music videos. Our code is available at \href{https://github.com/joeljang/music2video}{this https URL}. 
\end{abstract}

\section{Introduction}\label{sec:introduction}
% This template includes all the information about formatting manuscripts for KAIST GCT634/AI613.
% Please follow these guidelines to give the final proceedings a uniform look.
% Most of the required formatting is achieved automatically by using the supplied style file (\LaTeX) or template (Word) how to integrate audio and text prompts
The rapid development in deep learning has brought the visual generative models to not only create realistic images but to creatively render fictional images. Generative adversarial network models such as BigGAN, StyleGAN, and VQGAN exert their creative ability from randomly generated latent codes \cite{biggan, stylegan1, stylegan2, vqgan}. Since a single latent code corresponds to a particular image, modification in the latent space of GAN naturally creates a different image. In order to find meaningful manipulative directions in latent space of GAN, unsupervised methods such as singular value decomposition on sampled latent codes has been used.

On the other hand, common representation space of language and vision \cite{lazaridou2015combining, lu2019vilbert, tan2019lxmert, chen2020uniter} allows iterative update on the latent code of GAN to create a intended image which is given in a text format. For illustration, a popular multi-modal representation, CLIP \cite{clip}, which can encode the image and text into the common representation space is widely adapted with generative models. Since CLIP is a pre-trained model that can measure similarity between image and text in a shared latent space, variety of tasks that employ both image and text relies on the representational ability of CLIP. CLIP encodes the generated image and text guidance into a common vector space and compute the cosine similarity between the image encoding and text encoding to determine if they are similar.

Recent work extends the representational ability of CLIP into audio using diffusion \cite{wav2clip}. The shared representation space of image, text, and audio enables the latent code of GAN to generate an intended image when both text and audio given. However, naively integrating the audio representation into the text-based generation process tends to impedes the visualization of text representation. The conflicting behavior of the two modalities is the major obstacle in automatic music video generation. Moreover, it is vital to consider the property of music. Since music consists of varying themes which change irregularly, the created music video should correspondingly change its theme accordingly which is hard to capture when the music is segmented with fixed time interval.

The proposed method of this paper has two contributions.
\begin{itemize}

\item{We propose a method to seamlessly integrate the audio guidance and text guidance to create a music video of high quality.}
\item{We also show how to determine the change of scenery in the video based on the musical statistics.}
\end{itemize}

\section{Related Works}

\subsection{Common Representation of Diverse Modalities}
Learning common representation of image and text using contrastive objective has gained great attention for its versatile usage in text-guided generation and manipulation of image. Multi-modal representation learning such as CLIP and its variants \cite{clip, wav2clip} consist of two separate encoder networks each for different modality. For a pair of positive multi-modal samples, contrastive learning framework attempts to minimize the distance between the different modalities while pushing away the unrelated samples. The pre-trained model could be used for comparing the representation of image, text and audio. Works that optimize image generating process generally minimize the distance between the generated image representation and text prompt representation.

\begin{figure*}[t]
    \centering
    \includegraphics[width=\textwidth]{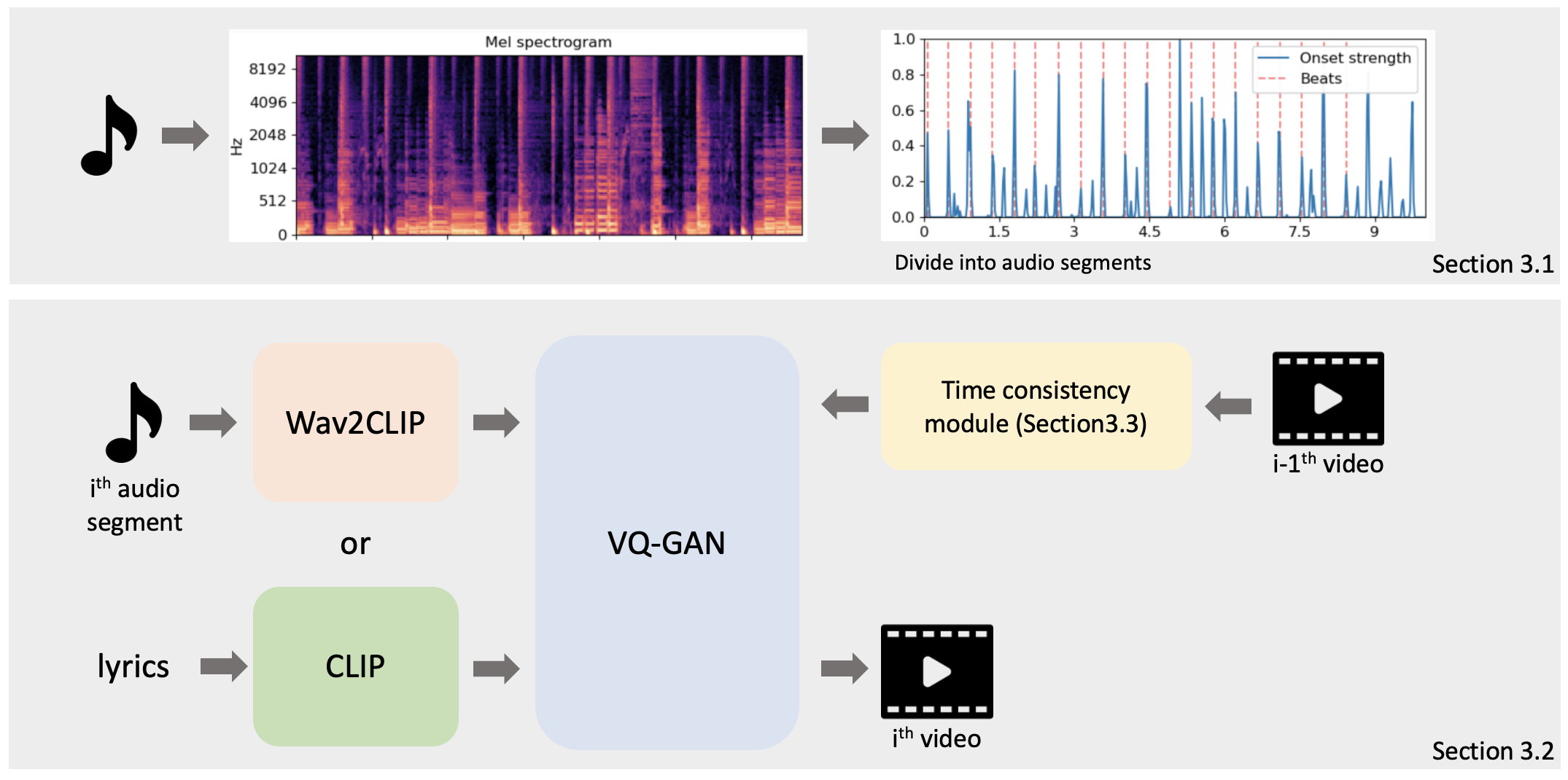}
    \caption{A graphical description of overall framework of music2video}
    \label{fig:framework}
\end{figure*}

\subsection{Vector Quantized GAN}
Convolutional neural network \cite{cnn} has been widely used for learning the representation of image since it captures the local patterns of a given image using kernels. However, CNN models fail to encode the long-range interaction between local features. In order to recognize the long-term dependency of high-level features using CNN, vector quantization \cite{vqvae, vqgan} summarizes the local features into a code book. The discretized code book enables transformer model to compute attention between the local features. Using the encoder network that maps the original image into the latent space, VQ-GAN additionally uses the vector quantization process to represent the latent space using code book. Using the discretized latent space, generator reconstructs the latent code into realistic images using adversarial loss between generator and discriminator.

\subsection{CLIP Based Image Generation}
Linear manipulation in the latent space of GAN corresponds to a different generated image. Such interpretable directions can found in both supervised~\cite{Shen_2020_CVPR, abdal2021styleflow} and unsupervised~\cite{NEURIPS2020_6fe43269, collins2020editing,  voynov2020unsupervised, Shen_2021_CVPR, wang2021aGANGeom, Jahanian*2020On, Patashnik_2021_ICCV} way. Recent methods~\cite{Patashnik_2021_ICCV, NEURIPS2020_6fe43269, collins2020editing} find such directions with the aid of multi-modal representation of CLIP. Text based image generation methods are built on off-shelf image generative models such as VQ-VAE \cite{ding2021cogview} and Big-GAN \cite{liu2021fusedream}.

\subsection{Visualizing Audio}
The extension of CLIP which is originally a visual-language model allows audio-based image generation. A method that is built on the pretrained Big-GAN model utilizes the intensity controlling property of the noise vector \cite{biggan} which adapts naturally to frame per second velocity based on the pitch of audio \cite{deep-music}. On the other hand, VQ-GAN is a unconditioned GAN which is capable of generating images out of ImageNET classes. Therefore, VQ-GAN based image generation could be widely adapted to express various themes \cite{vqgan-clip}.

\begin{figure*}[!t]
\begin{subfigure}{.23\textwidth}
  \centering
  % include first image
  \includegraphics[width=\textwidth]{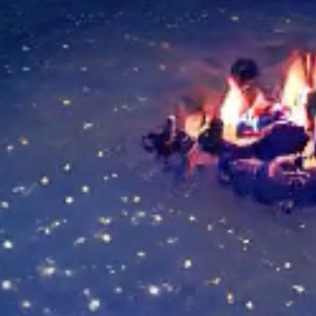}  
  \caption{L1 Loss: frame 1 - Campfire under sky full of stars}
  \label{fig:sub-first}
\end{subfigure}
\hfill
\begin{subfigure}{.23\textwidth}
  \centering
  % include second image
  \includegraphics[width=\textwidth]{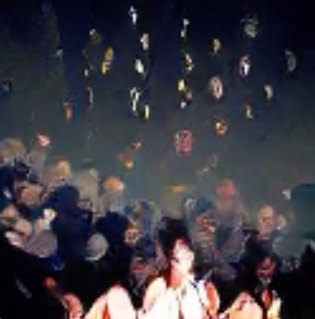}  
  \caption{L1 Loss: frame 2 - Crowd of people laughing}
  \label{fig:sub-second}
\end{subfigure}
\hfill
\begin{subfigure}{.23\textwidth}
  \centering
  % include third image
  \includegraphics[width=\textwidth]{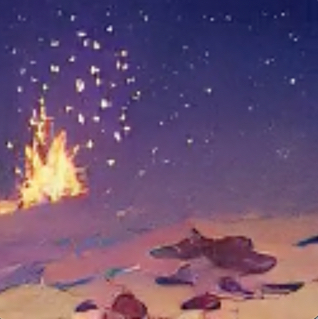}  
  \caption{Add prompts: frame 1 - Campfire under sky full of stars}
  \label{fig:sub-third}
\end{subfigure}
\hfill
\begin{subfigure}{.23\textwidth}
  \centering
  % include fourth image
  \includegraphics[width=\textwidth]{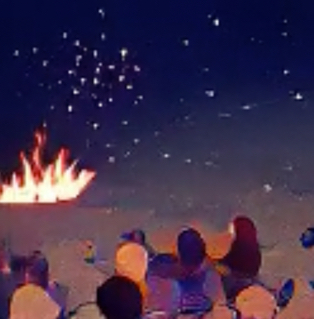}  
  \caption{Add prompts: frame 2 - Crowd of people laughing}
  \label{fig:sub-fourth}
\end{subfigure}
    \caption{Snapshots of a video sample with the corresponding text prompts.}
    \label{fig:time consistency}
\end{figure*}

\section{Our Framework}
We discuss how to automatically find a time stamp where scene transition should take place in Section \ref{sec:method1}. Based on the variable length segments of music, we demonstrate how to fuse the distinct guidance of audio and text domain to create sequence of images in Section \ref{sec:method2}. Based on the created sequence of images, we generate a video that maintains context from previous frame which is shown in detail in Section \ref{sec:method3}. The flow of overall framework is described in Figure \ref{fig:framework}.

\subsection{Video Scenery Transition based on Music}  \label{sec:method1}

In order to create a music video which changes the scenery accordingly with the music, we determine the time stamp where the musical statistics change drastically based on the strength of onset \cite{ellis2007beat}. We use the measured beat of music to split the music into dynamic intervals. The music used for demonstration is `Be my Weasel' \cite{yannic} composed by Yannic Kilcher. The music is divided into roughly 500 segments whose average time interval is 0.6 seconds. 

Given the segments of music, we determine frame-per-second based on the average intensity of the music segment. The given music is transformed into mel-spectrogram which shows intensity of each frequency in decibel for a given time interval. We compute average intensity of each segment then regularize the intensity of each segment to range between 0 and 1. The regularized intensity value corresponds to frame-per-second. Since each iteration of VQ-GAN generation process creates one image, frame-per-second is equivalent to the repeated number of image generation and latent code optimization process. 

\subsection{Iterative Audio and Text Guided Optimization} \label{sec:method2}
The naive approach of iteratively optimizing the generated image with audio and text fails to produce quality images. In this section we present details of optimizing generation process of VQ-GAN using both audio and text prompts. Given a segmented sections of music as illustrated in Section \ref{sec:method1}, we provide a single guidance for multiple frames in a single section. Since multiple frames correspond to the same number of iterations, each iteration will have the same audio or text guidance for an extended period. To illustrate further, each section is given an segmented audio prompt encoded into CLIP vector and only if the section consists of the time stamp given with the lyric, then the lyric encoded into CLIP vector is given instead. Unlike the iterative approach which alternates audio and text guidance for every iteration, this approach successfully represent both modalities of music and lyrics.

\subsection{Time Consistency in Video Generation} \label{sec:method3}
Since every frame in a video should be consistent with the previous frame, we explain how to maintain time consistency. We use two methods, L1 regularization of latent vectors and addition of guiding prompt vectors.

The first approach we experiment is to regularize the latent vector of VQ-GAN at iteration $t$ which is denoted as $z_t$ to have minimized l1 norm with the previous latent vector as in Equation \eqref{l1 norm}.

\begin{equation}\label{l1 norm}
min ||z_t - z_{t-1}||
\end{equation}

This method fails to show consistency of frames therefore we use the second approach Equation \eqref{add} adding the two distinct CLIP prompts for different frames. Providing a new guidance of adding the two CLIP vectors showed consistency in frames such as remaining objects from the previous frame as shown in Figure \ref{fig:time consistency}.

\begin{equation}\label{add}
\frac{CLIP(p_{t-1})+ CLIP(p_t)}{2}
\end{equation}

\begin{figure}[h]
\centering
\begin{subfigure}{.23\textwidth}
  \centering
  % include first image
  \includegraphics[width=\textwidth]{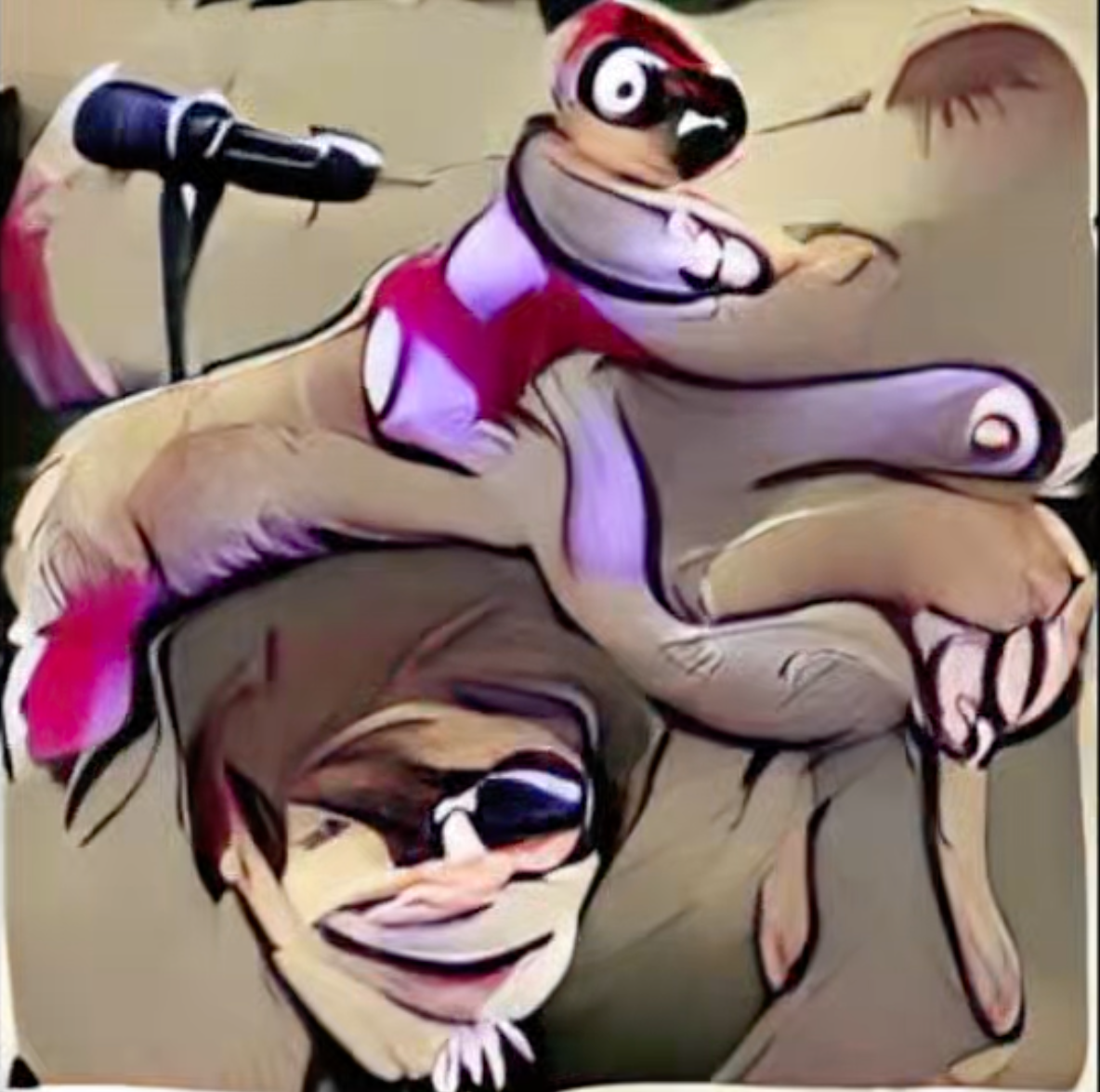}  
  \caption{be my weasel}
  \label{fig:sub-first}
\end{subfigure}
\begin{subfigure}{.23\textwidth}
  \centering
  % include second image
  \includegraphics[width=\textwidth]{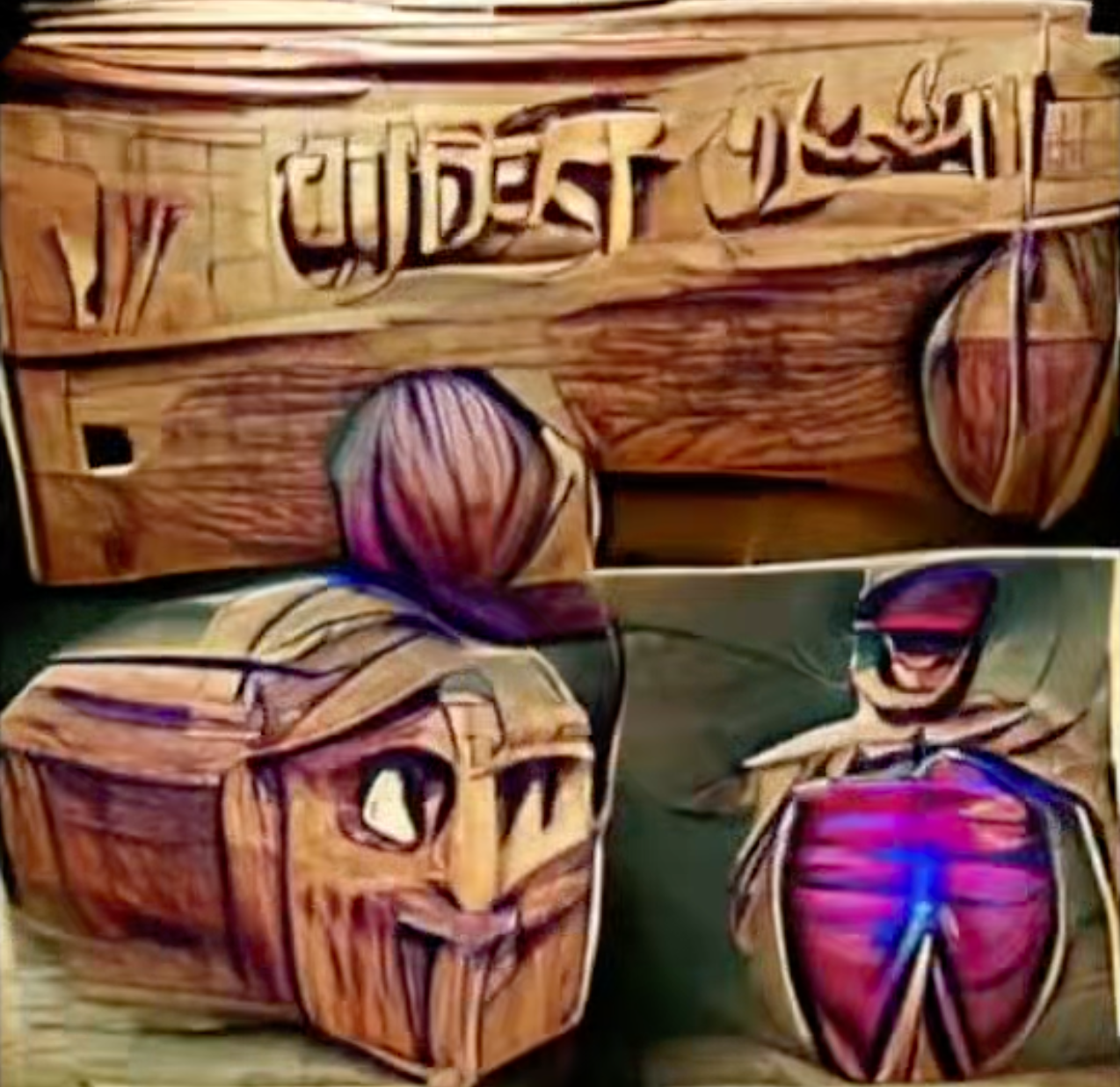}  
  \caption{what's inside this wooden chest}
  \label{fig:sub-second}
\end{subfigure}

\begin{subfigure}{.23\textwidth}
  \centering
  % include third image
  \includegraphics[width=\textwidth]{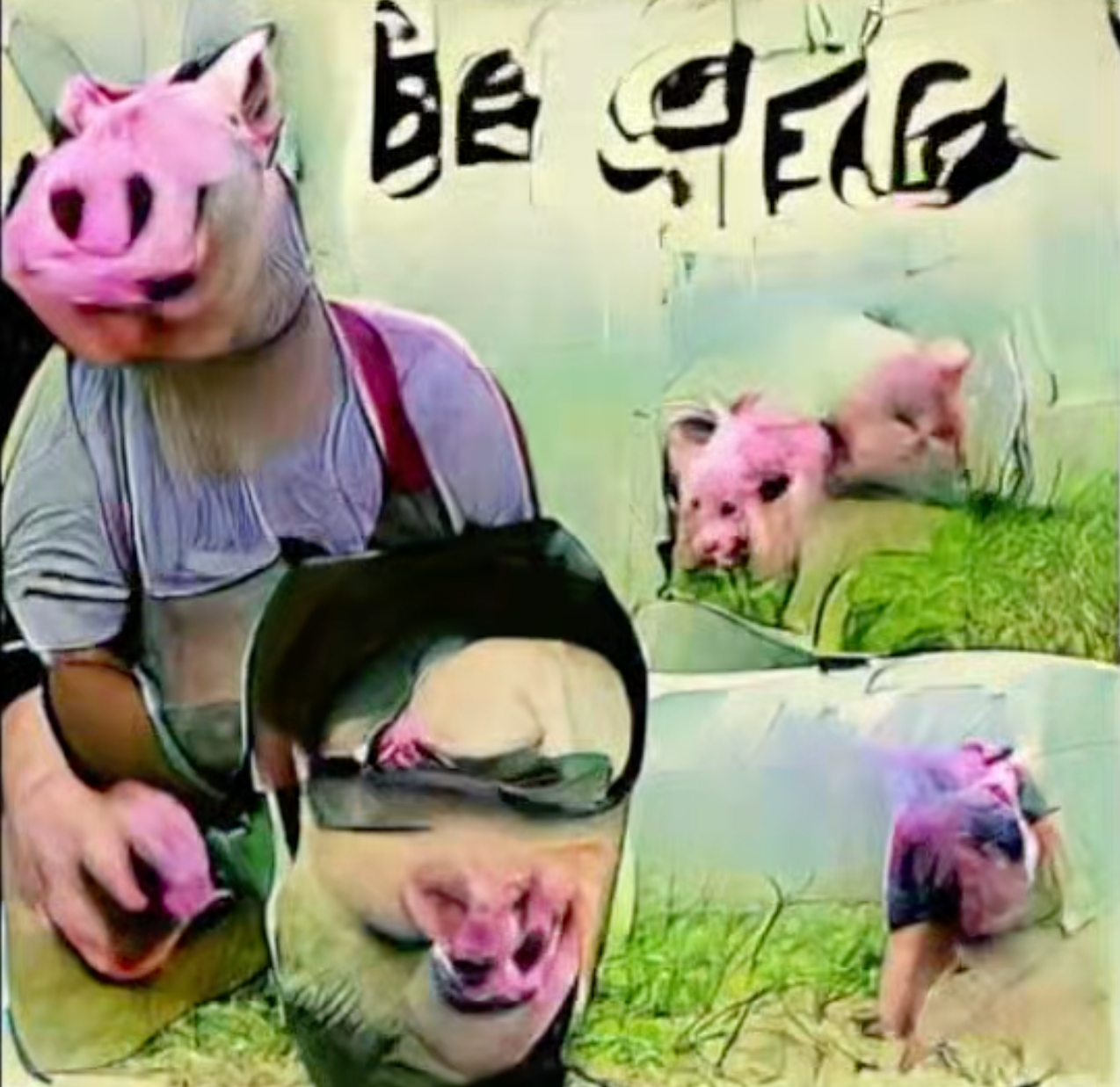}  
  \caption{be my pig}
  \label{fig:sub-third}
\end{subfigure}
\begin{subfigure}{.23\textwidth}
  \centering
  % include fourth image
  \includegraphics[width=\textwidth]{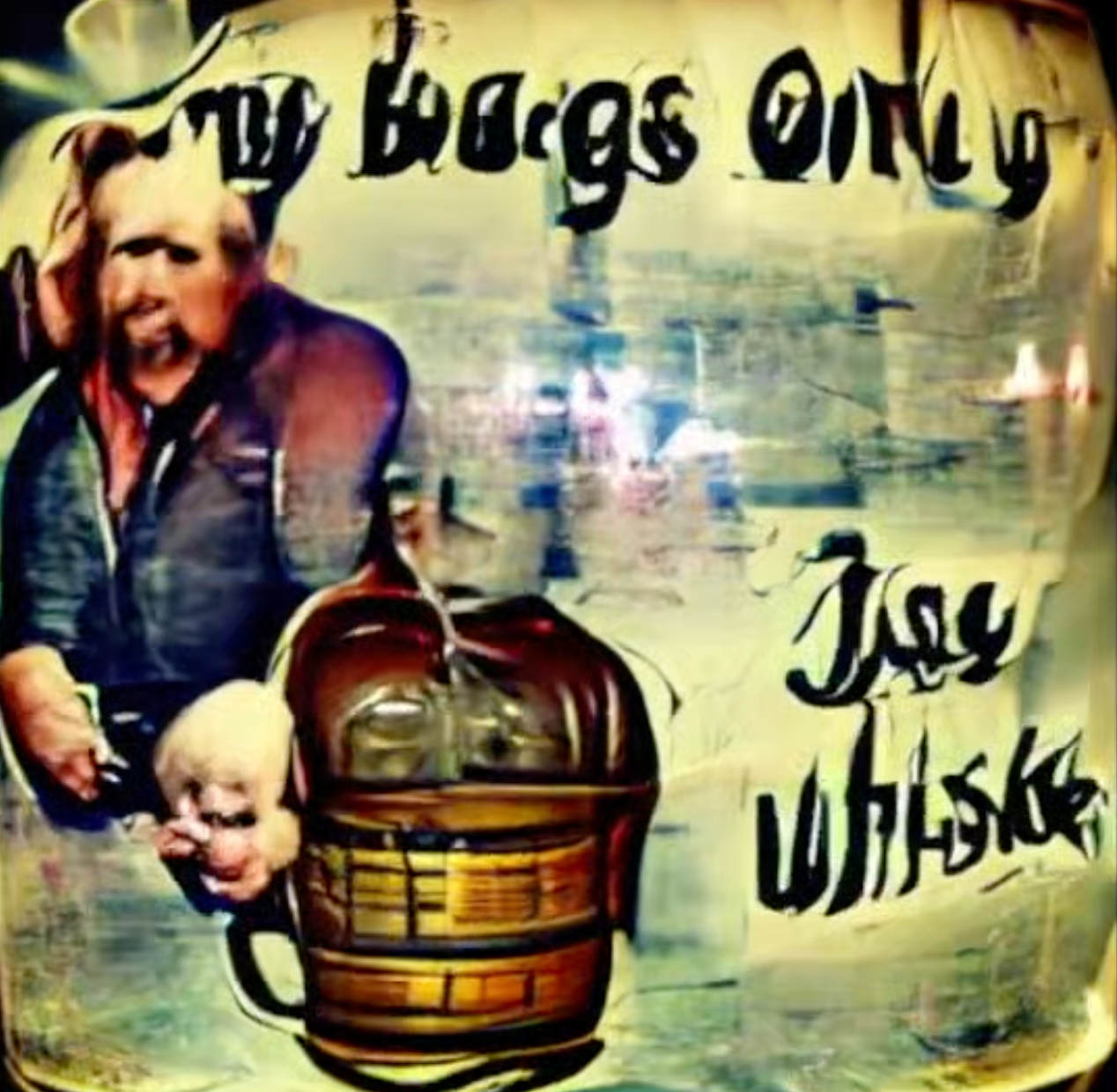}  
  \caption{and bring them all to my whiskey jug}
  \label{fig:sub-fourth}
\end{subfigure}
    \caption{Snapshots of a music video sample with the corresponding lyrics.}
    \label{fig:fig}
\end{figure}

% \subsection{Page Numbering, Headers and Footers}

% Third level headings are in Times 10pt italic, flush left,
% with 1/2 line of space above the section head, and 1/2 space below it.
% The first letter of each significant word is capitalized.

% \section{Footnotes and Figures}

% \subsection{Footnotes}

% Indicate footnotes with a number in the text.\footnote{This is a footnote.}
% Use 8pt type for footnotes. Place the footnotes at the bottom of the page on which they appear.
% Precede the footnote with a 0.5pt horizontal rule.

% \subsection{Figures, Tables and Captions}

% All artwork must be centered, neat, clean, and legible.
% All lines should be very dark for purposes of reproduction and art work should not be hand-drawn.
% The proceedings are not in color, and therefore all figures must make sense in black-and-white form.
% Figure and table numbers and captions always appear below the figure.
% Leave 1 line space between the figure or table and the caption.
% Each figure or table is numbered consecutively. Captions should be Times 10pt.
% Place tables/figures in text as close to the reference as possible.
% References to tables and figures should be capitalized, for example:
% see \figref{fig:example} and \tabref{tab:example}.
% Figures and tables may extend across both columns to a maximum width of 17.2cm.

% For bibtex users:
\bibliography{ms.bib}

\end{document}